# LINAC

*D. Alesini*

Laboratori Nazionali di Frascati dell'INFN, Frascati, Italy

**Abstract**

A linac (linear accelerator) is a system that allows to accelerate charged particles through a linear trajectory by electromagnetic fields. This kind of accelerator finds several applications in fundamental research and industry. The main devices used to accelerate the particle beam will be introduced in the first part of the paper, while in the second part, the fundamentals of the longitudinal and transverse beam dynamics will be highlighted. A short paragraph is finally dedicated to radiofrequency quadrupoles (RFQ).

## 1    Introduction

A linear accelerator (linac) is a device that allows accelerating charged particles (electrons, protons, ions) along a straight trajectory [1,2]. The main advantage of linacs is their capability to produce high-energy, high-intensity charged particle beams of excellent quality in terms of beam emittance and energy spread. These devices find applications in different fields such as research, healthcare and industry [3-8].

The acceleration can be obtained with constant (DC) or time-varying electric fields. In the second case, we can have two types of linacs: radio frequency (rf) and induction. In this proceeding we will refer to rf linacs in which the particle acceleration is obtained by electromagnetic fields confined in resonant cavities fed by sinusoidal time-varying power sources. The design and the structure of a linac depend on the type of accelerated particles (e.g. electrons, protons or ions) and on the required final beam parameters in terms of energy, energy spread, emittance and current. In the design and in the choice of the technology, several constraints have also to be taken into account such as cost, available footprint and power consumption.

The main advantages [9] of linacs with respect to other possible accelerators (as synchrotrons or cyclotrons) are the fact that they can handle high peak current beams, they can run with high duty-cycle and they exhibit low synchrotron radiation losses (in the case of light particle acceleration as electron or positrons). On the contrary, the main drawback is the fact that they require a large number of cavities to reach a desired energy, since the beam passes only once in the accelerating structures. Moreover, synchrotron radiation damping (in the case of light particles) cannot be used to reduce the beam emittance unless to adopt complicated schemes.

Linacs are mainly used in fundamental physics research as injectors for synchrotrons and storage rings, free electron lasers and injectors for colliders. They also find a huge number of medical and industrial applications for cancer therapy, x rays generation, material treatment, food irradiation and ion implantation.

A linac conceptual scheme, where its technology complexity can be also assessed, is shown in Fig. 1. Particles are generated and pre-accelerated in the injector, and then accelerated by accelerating structures and focused by magnetic elements (quadrupoles and solenoids). Beam trajectory and dimensions along the linac are measured by different types of diagnostic devices (striplines, cavity beam position monitors and beam screens) depending on the particle beam and energy. In the figure are reported also the control, cooling and vacuum systems, the rf distribution and rf power sources.

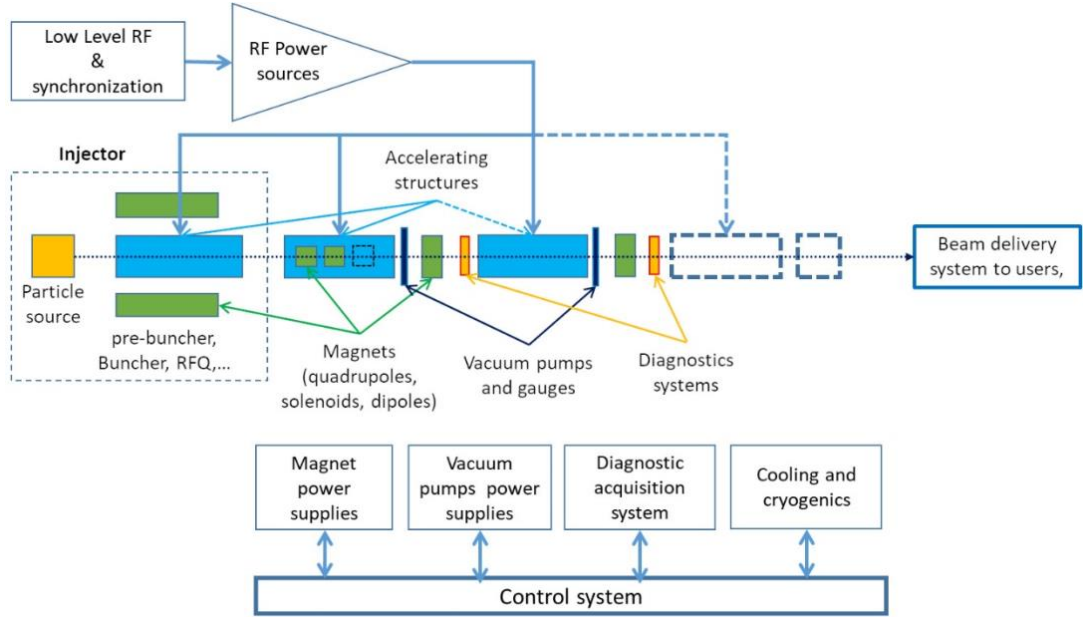

**Fig.1:** Schematic layout of a linac

## 2     Acceleration process: particle velocity and energy

The first historical linear accelerator was conceived by the Nobel prize Wilhelm Conrad Röntgen (1901). It consisted in a vacuum tube containing a cathode connected to the negative pole of a DC voltage generator of few kV. Electrons emitted by the heated cathode (by thermionic emission) were accelerated while flowing to another electrode connected to the positive generator pole (anode). Collisions between the energetic electrons and the anode produced x-rays and gave the possibility to do the first radiography.

In modern electrostatic accelerators high voltage is shared between a set of electrodes creating an accelerating field between them. This type of accelerator is better known as the Cockcroft-Walton accelerator [10]. Its main limitation in term of achievable energy, is due to the fact that all partial accelerating voltages add up at some point and insulation problems or discharges occur, thus limiting the maximum achievable voltage to a few tens of MV.

Electrostatic accelerators are still used for several applications as x ray generation, material analysis, or ion implantation. The simple scheme of a DC acceleration process between two electrodes is reported in Fig. 2(a).

Considering the energy-momentum relation and the Lorentz force we can easily write the following relations:

$$E^2 = E_0^2 + p^2c^2 \Rightarrow 2EdE = 2pdpc^2 \Rightarrow dE = v\frac{mc^2}{E}dp \Rightarrow dE = vdp \tag{1}$$

$$\frac{dp}{dt} = qE_z \underset{z=vt}{\Rightarrow} v\frac{dp}{dz} = qE_z \Rightarrow \frac{dE}{dz} = qE_z \quad \left(\text{and also} \frac{dW}{dz} = qE_z\right), \tag{2}$$

where $E_0$ $(=m_0c^2)$ is the particle rest energy, $E$ is the total energy, $m_0$ is the rest mass, $m$ is the relativistic mass, $v$ is the particle velocity, $p$ $(=mv)$ is the particle momentum, $W=E-E_0$ is the kinetic energy and $E_z$ is the accelerating field. In Eq. (2) $dE/dz$ is the rate of energy gain per unit length and it is proportional to the accelerating field $E_z$.

Integrating Eq. (2) on the accelerating gap we obtain the energy gain per electrode pair $\Delta E$:

$$\Delta E = \int_{gap} \frac{dE}{dz} dz = \int_{gap} qE_z dz \Rightarrow \Delta E = q\Delta V \quad . \tag{3}$$

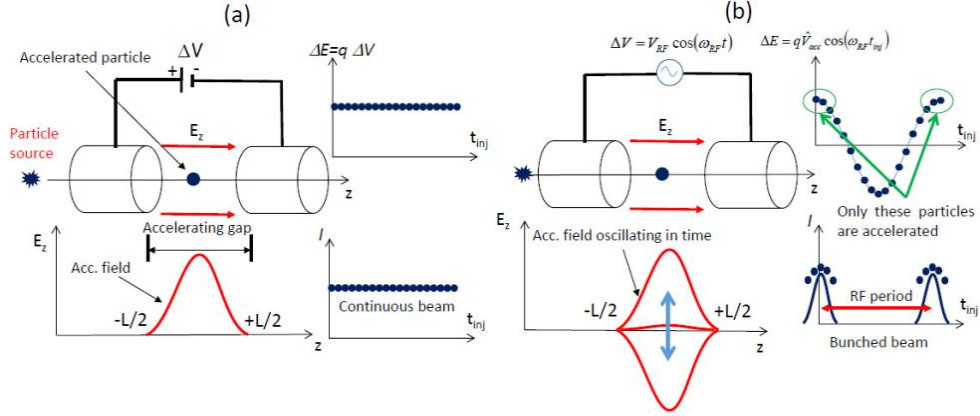

**Fig.2:** Scheme of acceleration processes between two electrodes: (a) electrostatic acceleration; (b) rf acceleration.

If now we consider the relativistic factors $\beta=v/c$ and $\gamma=E/E_0$ we can write:

$$\beta = \sqrt{1-\frac{1}{\gamma^2}} = \sqrt{1-\left(\frac{E_0}{E}\right)^2} = \sqrt{1-\left(\frac{E_0}{E_0+W}\right)^2} \quad . \tag{4}$$

The behavior of $\beta$ and $\gamma$ as a function of the kinetic energy is given in Fig. 3 for an electron and a proton. From the plot it is evident that light particles (electrons) are practically fully relativistic ($\beta \cong 1$, $\gamma \gg 1$) at relatively low energies (above ~10 MeV) and reach a constant velocity (~$c$). Thus, for this type of particles, the acceleration process occurs at constant velocity. On the contrary heavy particles (protons and ions) are typically weakly relativistic and reach a constant velocity only at very high energies. This means that their velocity changes during the acceleration process. As illustrated in the following, the possible velocity change during acceleration generates important differences in the accelerating structures design and beam dynamics between light and heavy particles.

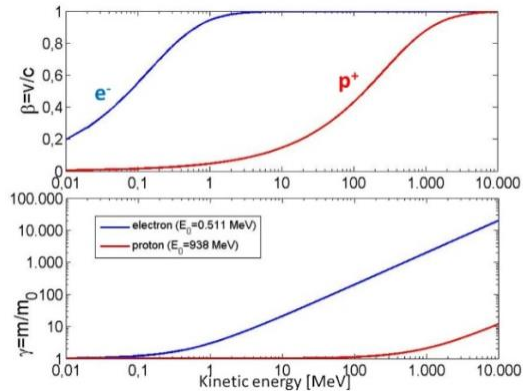

**Fig. 3:** $\beta$ (top) and $\gamma$ (bottom) as a function of the kinetic energy for an electron (blue) and a proton (red).

# 3 Radiofrequency accelerators: the Wideröe linac

In the late 1920's propositions were made by Ising (1924) and implemented by Wideröe (1927) to overcome the limitation of electrostatic devices in terms of reachable energy [11-13]. The proposed scheme is illustrated in Fig. 4. An AC voltage generator feeds, alternately, a series of electrodes in such a way that particles do not experience any force while travelling inside the tubes (Drift Tubes) while are accelerated across the gaps. In particular, this last statement is true if the drift tube length (or, equivalently, the distances between the centers of the accelerating gaps $L_n$) increases with the particle velocity during the acceleration, such that the time of flight between gaps is kept constant and equal to half of the rf period. If this condition is satisfied, the particles are subject to a synchronous accelerating voltage and experience an energy gain of $\Delta E = q\Delta V$ at each gap.

In this type of structures, called Drift Tube Linac (DTL), a single rf generator can be used, in principle, to indefinitely accelerate the beam, avoiding the breakdown limitations that affect the electrostatic accelerators.

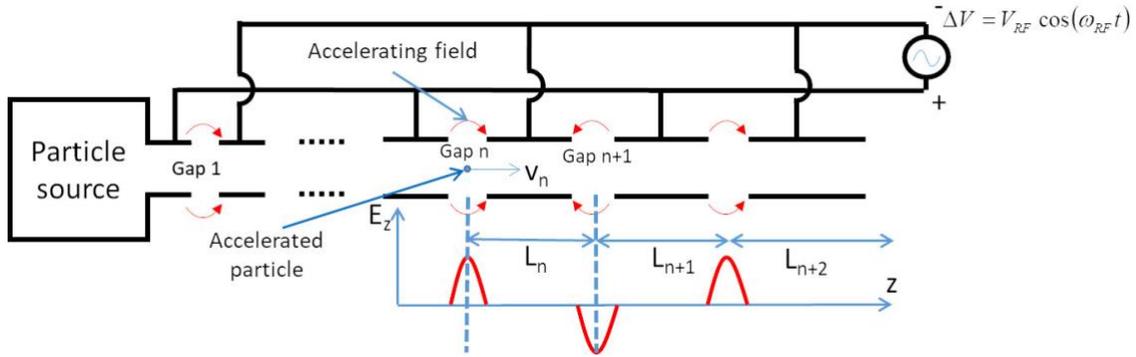

**Fig. 4:** Conceptual scheme of a Wideröe linac.

The Wideröe linac is the first rf linac. As illustrated in Fig. 1(b), a rf acceleration process does not allow to accelerate a continuous particle beam but the beam needs to be "bunched" with a distance between bunches exactly equal to the rf period.

Let us now consider the acceleration between a pair of two electrodes. The behavior of the accelerating electric field and voltage are given by the following relations:

$$E_z(z,t) = E_{RF}(z)\cos(\omega_{RF} t)$$

$$\Delta V = V_{RF}\cos(\omega_{RF} t) \qquad \omega_{RF} = 2\pi f_{RF} = \frac{2\pi}{T_{RF}} \qquad V_{RF} = \int_{gap} E_{RF}(z)dz \quad . \tag{5}$$

The energy gain per electrode (supposing a symmetric accelerating field with respect to $z=0$) is given by:

$$\Delta E = q\int_{gap} E_z(z,t)\Big|_{\substack{seen\\by\\particle}} dz = q\int_{-L/2}^{+L/2} E_{RF}(z)\cos\left(\omega_{RF}\frac{z}{v} + \phi_{inj}\right)dz = q\underbrace{\int_{gap} E_{RF}(z)dz}_{V_{RF}} \underbrace{\frac{\int_{gap} E_{RF}(z)\cos\left(\omega_{RF}\frac{z}{v}\right)dz}{\int_{gap} E_{RF}(z)dz}}_{T} \cos(\phi_{inj}) = q\underbrace{V_{RF}T}_{\hat{V}_{acc}}\cos(\phi_{inj})$$

(6)

where $\phi_{inj}$ is the injection phase of a generic particle. $T$ is called the "transit time factor" and is always less than 1. It takes into account the fact that the rf voltage is oscillating in time while the beam is traversing the gap and that the effective peak accelerating voltage ($\hat{V}_{acc}$) is in fact equal to the rf voltage ($V_{RF}$) multiplied by this factor. $E_{acc} = V_{acc}/L [V/m]$ is the average accelerating field seen by the particle.

If we consider a Wideröe structure, the energy gain, at each gap, is equal to $\Delta E_n = qV_{acc}$ while the particle velocity increases accordingly to Eq. (4). In order to be synchronous with the accelerating field at the center of each gap, the time of flight ($t_n$) between gaps is given by:

$$t_n = \frac{L_n}{\bar{v}_n} = \frac{T_{RF}}{2} \Rightarrow L_n = \frac{1}{2}\bar{v}_n T_{RF} = \frac{1}{2}\bar{\beta}_n \underbrace{cT_{RF}}_{\lambda_{RF}} \quad , \tag{7}$$

where $T_{RF}$ is the generator rf period, $\bar{v}_n$ is the average particle velocity between the gap $n$ and $n+1$ and $\lambda_{RF}$ is the rf wavelength. Then, the distance between the centers of two consecutive gaps has to be increased as follows:

$$L_n = \frac{1}{2}\bar{\beta}_n \lambda_{RF} \quad . \tag{8}$$

The energy gain per unit length is then given by:

$$\frac{\Delta E}{\Delta L} = \frac{qV_{acc}}{L_n} = \frac{2qV_{acc}}{\lambda_{RF}\bar{\beta}_n} \quad . \tag{9}$$

## 4 Rf cavities

There are two important consequences of Eq. (9) obtained in the previous paragraph. First of all, the condition $L_n \ll \lambda_{RF}$ (necessary to model the tube as an equipotential region) requires $\bar{\beta}_n \ll 1$. This implies that the Wideröe acceleration technique cannot be applied to relativistic particles. Moreover, when particles velocity increases the drift tube gets longer, reducing the acceleration efficiency (energy gain per unit length $\Delta E/\Delta L$). The average accelerating gradient (defined as $E_{acc} = V_{acc}/L_n$) increase pushes towards the use of small $\lambda_{RF}$ (high frequencies), but the concept of equipotential drift tubes cannot be applied at small $\lambda_{RF}$ and the power lost by free space radiation increases proportionally to the rf frequency.

All previous considerations combined with the fact that high frequency, high power rf sources became available only after the 2$^{nd}$ world war (thanks to the radar technology applied to military purposes), pushed to develop more efficient accelerating structures than simple drift tubes. In order to avoid electromagnetic (e.m.) radiation processes and to allow the use of high frequency sources, the accelerating system had to be enclosed in a metallic volume: a rf cavity. In a rf cavity the e.m. field has a particular spatial configuration (resonant mode) whose components, including the accelerating field $E_z$, oscillates at a specific frequency (called resonant frequency) characteristic of the mode. The mode is excited by the rf generator that is coupled to the cavity through waveguides, coaxial cable antennas or loops. The resonant modes are called Standing Wave (SW) modes since they have a spatial fixed configuration that oscillates in time. The spatial and temporal field profiles in a cavity have to be computed (analytically or numerically) by solving the Maxwell equations with the proper boundary conditions [14,15].

For a SW cavity the accelerating field on the z-axis can be written as:

$$E_z(z,t) = E_{RF}(z)\cos\left(\underbrace{2\pi f_{RF}}_{\omega_{RF}} t + \phi\right) = \text{Real}\left[\tilde{E}_z(z)e^{j\omega_{RF}t}\right] \quad , \tag{10}$$

where $f_{RF}$ is the excitation frequency of the generator equal to (or close to) the resonant frequency of the cavity, $\omega_{RF}$ is the angular excitation frequency, $E_{RF}(z)$ (or the phasor $\tilde{E}_z$) is a real (complex) function related to the spatial configuration of the mode.

Rf linacs use different type of cavities depending on the species and energy range of particles to be accelerated, as described in the following paragraphs.

## 5 Alvarez structures

The Alvarez structures [16], reported in Fig. 5, can be described as special DTLs in which the electrodes are part of a resonant macrostructure. They work in the so-called "0-mode", since the accelerating field at a given time has the same phase in each gap. The Wideroe structures, on the contrary, work in the "π-mode", i.e. the accelerating field in consecutive gaps has opposite sign. They are used for protons and ions in the range of *β=0.05-0.5*. They typically operate at $f_{RF}$=*50-400 MHz*, $\lambda_{RF}$=*6-0.7 m* in the *1-100 MeV* energy range. Usually, they are simply called DTL (instead of the Wideroe structures that are not used anymore).

As for the Wideroe structure, also in the Alvarez linac the accelerating field is concentrated between gaps and the beam crosses the drift tubes when the electric field is decelerating. The drift tubes are suspended by stems and quadrupole magnets (for transverse focusing) can fit inside them.

To maintain the synchronism between the beam and the accelerating field at each gap, the distance between the accelerating gaps $L_n$ has to be varied according to the formula:

$$L_n = \overline{\beta}_n \lambda_{RF} \tag{11}$$

To maintain this synchronism, in principle, both the gap and the drift tube length can be varied. A generally applied criterion is to maintain a constant transit time factor according to Eq. (6).

Some examples of Alvarez structures can be found at CERN linac 4 [17] operating at *352 MHz* frequency with 3 resonators tanks of ~500 mm diameter, 19 m long, 120 Drift Tubes, that accelerate protons from 3 MeV to 50 MeV (β=0.08 to 0.31). In this case the distance between gaps varies from 68mm to 264mm.

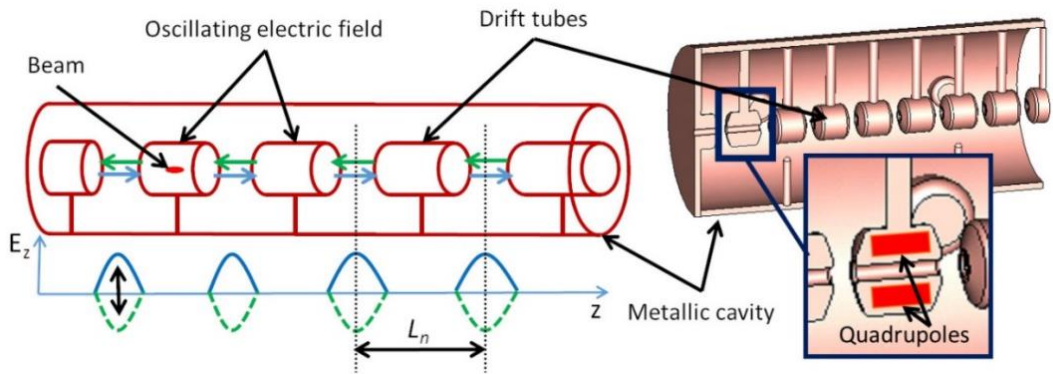

**Fig. 5:** Schematic views of the Alvarez structure (Drift Tube Linac).

## 6 High β cavities: cylindrical structures

When the β of the particles increases (>0.5) one has to use higher rf frequencies (>400-500 MHz) to increase the accelerating gradient per unit length. It is possible to demonstrate that DTL structures become less efficient (effective accelerating voltage per unit length for a given input power) for such values of β.

In this range of frequencies, cylindrical single or multiple cavities working in the $TM_{010}$-like mode are used. For a pure cylindrical structure (usually called a "pillbox cavity") the first accelerating mode (i.e. the mode with non-zero longitudinal electric field on axis) is the $TM_{010}$ mode. It has a well-known analytical solution (obtained solving Maxwell's equations), and its spatial configuration in the case of a pure cylindrical cavity is given in Fig. 6(a). For this mode the electric field has only a longitudinal component, while the magnetic one is purely azimuthal. The corresponding complex phasors are given by [14]:

$$\begin{cases} \tilde{E}_z = A J_0\left(p_{01}\frac{r}{a}\right) \\ \tilde{H}_\theta = -jA\frac{1}{Z_0} J_0'\left(p_{01}\frac{r}{a}\right) \end{cases}, \quad (12)$$

where $a$ is the cavity radius, $A$ is the mode amplitude and $p_{01}$ (= 2.405) is the first zero of the Bessel function $J_0$. The resonant frequency of this mode is given by: $f_{res} = \dfrac{p_{01} c}{2\pi a}$.

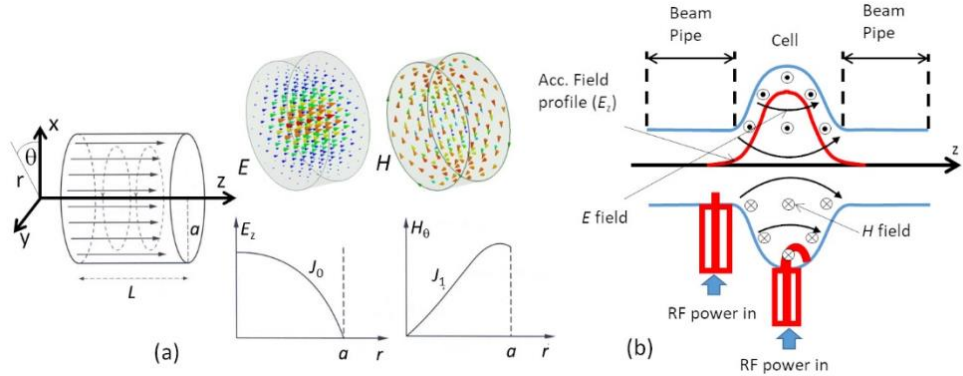

**Fig. 6:** (a) Ideal pillbox cavity and e.m. field configuration; (b) Sketch of a real cavity operating in the $TM_{010}$-like mode with two type of coaxial couplers (antenna and loop).

The geometry of real cylindrical cavities is somewhat different to that of a pillbox. In fact, one has to also consider the perturbation introduced by the beam pipe, the power couplers to rf generators and any antenna or pick-up used to monitor the accelerating field inside the cavity. For this reason, the actual accelerating mode is called the $TM_{010}$-like mode. Real cavities and their couplers to the rf generators are designed using numerical codes that solve the Maxwell equations with the proper boundary conditions. A sketch of a cavity fed through a loop or a coaxial probe to an external generator is given in Fig. 6(b), where there are also reported the electric and magnetic field lines and the longitudinal electric field profile on axis. Details of the coupler design can be found in [18].

## 6.1 SW cavity parameters: shunt impedance and quality factor

For a SW cavity the first figure of merit is the shunt impedance defined by [1,2,14,19]:

$$R = \frac{\hat{V}_{acc}^2}{P_{diss}} \quad [\Omega], \quad (13)$$

where $\hat{V}_{acc}$ is the accelerating voltage for a given dissipated power into the cavity ($P_{diss}$). This parameter qualifies the efficiency of the cavity: the higher its value, the larger is the achievable accelerating voltage for a given dissipated power. As an example, at 1 GHz for a normal conducting (NC) copper cavity a typical shunt impedance of the order of 2 MΩ can be obtained, while a superconducting (SC) cavity, at the same frequency, can reach values of the order of 1 TΩ, due to the extremely lower dissipated power. It is also useful to refer to the following quantity called shunt impedance per unit length:

$$r = \frac{R}{L} = \frac{(\hat{V}_{acc}/L)^2}{P_{diss}/L} = \frac{\hat{E}_{acc}^2}{p_{diss}} \quad [\Omega/m], \quad (14)$$

where $L$ is the cavity length, $\hat{E}_{acc}$ the average accelerating field and $p_{diss}$ the dissipated power per unit length.

The quality factor of the accelerating mode is then defined by the ratio of the cavity stored energy ($W$) and the dissipated power on the cavity walls:

$$Q_0 = \omega_{RF} \frac{W}{P_{diss}}. \quad (15)$$

For a NC cavity operating at 1 GHz the quality factor is of the order of $10^4$ while, for an SC cavity, values of the order $10^9$ to $10^{10}$ can be achieved.

It can be easily demonstrated that the ratio $R/Q$ is a pure geometric factor and it does not depend upon the cavity wall conductivity or operating frequency. This is the reason why it is always taken as a geometric design qualification parameter. The $R/Q$ of a single cell is of the order of 100.

## 6.2 Multi-cell SW cavities

In linacs, rf cavities are used in systems of multi-cavities. In a multi-cell structure, there is one input coupler that feeds a system of coupled cavities as sketched in Fig. 8. The field of adjacent cells is coupled through the cell irises (and/or through properly designed coupling slots). It is quite easy to demonstrate that, for such structures the shunt impedance is $N$ times the impedance of a single cavity. Moreover, with one source, it is possible to feed a set of cavities with a simplification of the power distribution system layout. On the other hand, the fabrication of multi-cell structures is more complicated than single-cell cavities.

## 6.3 π-mode structures

The $N$-cell structure behaves like a system of $N$ coupled oscillators with $N$ coupled multi-cell resonant modes. As an example, the field configuration of a two-cell resonator is shown in Fig. 7. The mode in which the two cells oscillate with the same phase is called 0 mode, while the one with 180° phase shift is called π-mode. It is quite easy to verify that the most efficient configuration (generally used for acceleration) is the π-mode, which is shown in Fig. 8 for a system of five cells. In this system, in order to have a synchronous acceleration in each cell, the distance ($d$) between the centre of two adjacent cells has to be $d = v/(2f_{RF})$ where $v$ is the particle velocity. As sketched in Fig. 8 this allows to synchronize the beam passage in each cell with the accelerating field, granting a continuous acceleration process.

To maintain the synchronism, for ions and protons the cell length has to be increased and the linac will be made of a sequence of different accelerating structures matched to the ion/proton velocity. For electrons, after the injector, $d=\lambda_{RF}/2$ and the linac is made by a series of identical accelerating structures.

Examples of this type of system are the LINAC 4 (CERN) PIMS (PI Mode Structure) that operates at $f_{RF}=352$ MHz with $\beta>0.4$ [17] while, for electrons, the superconducting cavities of the European X-FEL that operate with modules of nine coupled cells at 1.3 GHz [20-21].

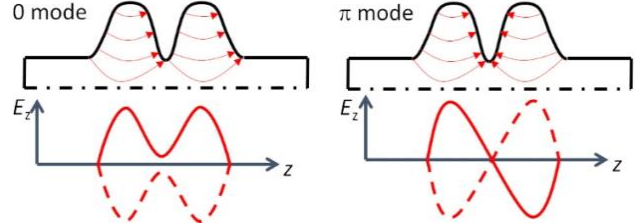

**Fig. 7:** Resonant modes in a system of two coupled cavities. The mode typically used for acceleration is the π-mode.

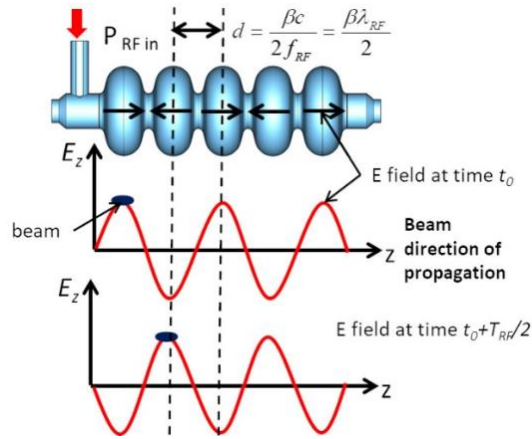

**Fig. 8:** Acceleration in multi-cell SW cavity operating on the π mode: the cell length is equal to $\beta\lambda_{RF}/2$ and the bunch in each cell is always synchronous with the electric field positive half-wave.

### 6.4   π/2 mode structures

It is possible to demonstrate that in a multi-cell system the different resonant modes are distributed in a curve called "dispersion curve". As an example, the case of a 9-cell structure is reported in Fig. 9. Each mode has a bandwidth proportional to the quality factor [19] and, over a certain number of coupled cavities, the overlap of the tails of adjacent modes can introduce problems for field equalization and structure operability. This limits the maximum number of multi-cell structures to around 10-15. Since the criticality of a working mode depend on the frequency separation between the working mode and the adjacent modes, the π/2 mode, from this point of view, is the most "stable" one. Unfortunately, for this mode it is possible to demonstrate that the accelerating field is zero every two cells. One possible solution to use this mode is to put off-axis the empty cells (called coupling cells) like in the system schematically represented in Fig. 10(a). In spite of this mechanical complication with respect to a π-mode system (see as example the mechanical drawing of Fig. 10(b)), this allows to increase the number of cells to more than 20 without problems.

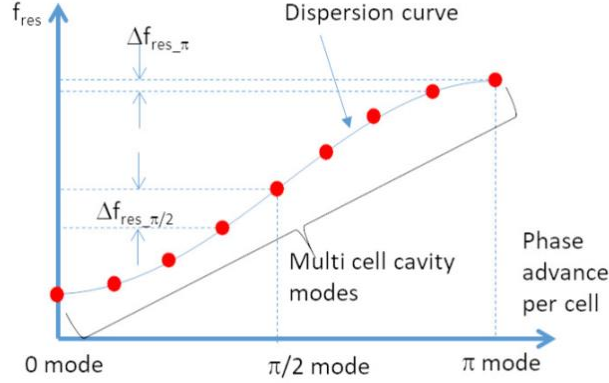

**Fig. 9:** Resonant modes distributed in the "dispersion curve" for a 9-cell multi-cell structure.

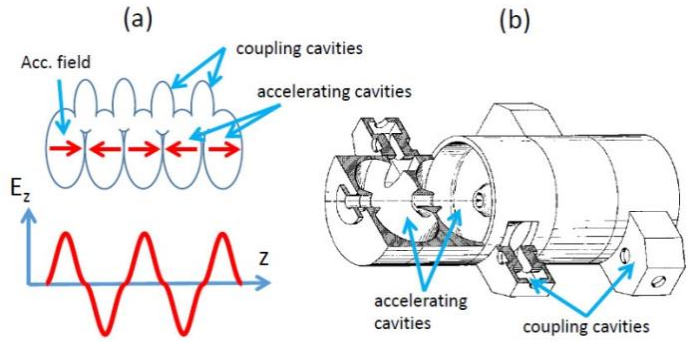

**Fig. 10:** (a) Sketch of a multi cell cavity of 5 cell operating on the π/2 mode; (b) Mechanical sketch of the cavity.

These types of structures are used both for electrons than protons. As example in the Spallation Neutron Source [22] there is the Coupled Cavity Linac (CCL) section with 4 modules, each containing 12 accelerator segments. It operates at 805 MHz and accelerates the proton beam from 87 to 186 MeV in a length of about 55 meters.

## 7  Travelling wave structures

There is another possibility to accelerate particles: using travelling wave (TW) structures. In TW structures an e.m. wave travels together with the beam in a special guide in which the phase velocity of the wave matches the particle velocity. In this case the beam absorbs energy from the wave and it is continuously accelerated.

Typically, these structures are used for electrons because for such particles the velocity can be assumed constant all along the structure and equal to *c*, while it would be difficult to modulate the phase velocity for heavy particles that change their velocity during acceleration.

In the simple case of an e.m. wave propagating into a constant cross-section waveguide, the phase velocity is always larger than the speed of light and thus the e.m. wave will never be synchronous with a particle beam. As example in a circular waveguide (Fig. 11(a)) the first propagating mode with $E_z \neq 0$ is the $TM_{01}$ mode, whose longitudinal electric field (neglecting the attenuation) can be expressed by the well know formula [23]:

$$E_z\big|_{TM_{01}} = E_0 J_0\left(\frac{p_{01}}{a} r\right) \cos(\omega_{RF} t - k_z^* z) \qquad k_z^* = \frac{1}{c}\sqrt{\omega_{RF}^2 - \omega_{cut}^2} \quad , \tag{16}$$

where *a* is the radius of the waveguide, $k_z^*$ is the propagation constant, $\omega_{cut}$ is the cut-off angular frequency of the waveguide equal to $\omega_{cut} = c p_{01}/a$. The corresponding phase velocity is given by:

$$v_{ph} = \frac{\omega_{RF}}{k_z^*} = \frac{c}{\sqrt{1-\omega_{cut}^2/\omega_{RF}^2}} \quad , \qquad (17)$$

which is always larger than *c*.

The behaviour of the propagation constant as a function of frequency is the well-known dispersion curve and is sketched in Fig. 11(a). It is important to remark that the phase velocity is not the velocity of the energy propagation into the structure, which, instead, is given by the group velocity ($v_g$):

$$v_g = \left.\frac{d\omega}{dk_z}\right|_{\omega=\omega_{RF}} = c\sqrt{1-\omega_{cut}^2/\omega_{RF}^2} \qquad (18)$$

and is always smaller than *c*.

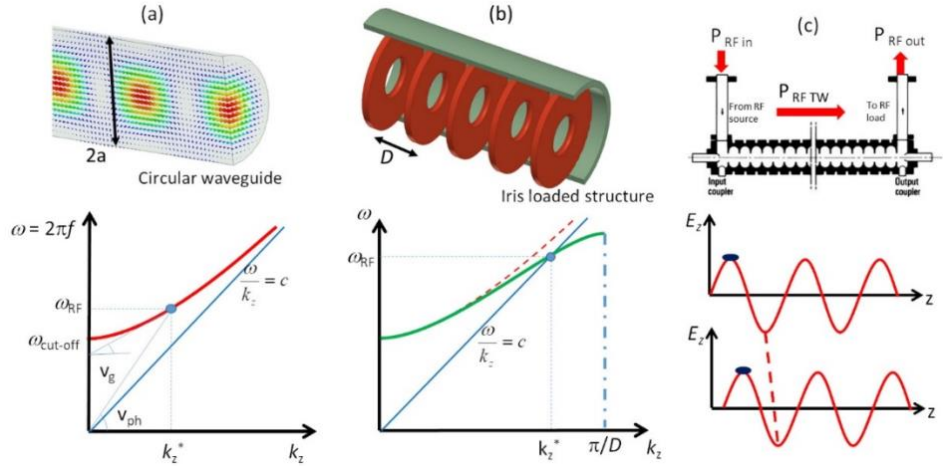

**Fig. 11:** (a) Circular waveguide example (top) and its dispersion curve (bottom); (b) Iris loaded waveguide model (top) and its typical dispersion curve (bottom); (c) TW structure with input and output couplers: the e.m. field travels together with the bunch, continuously transferring its energy to the particles.

In order to slow down the wave phase velocity, the structure through which the wave is travelling, is periodically loaded with irises. A sketch of an iris-loaded structure is given in Fig 11(b) and it can be designed to have the phase velocity equal to the speed of the particles allowing a net acceleration over large distances. The propagating field, in this case, is that of a special wave travelling within a spatial periodic profile ($TM_{01}$-like mode) and, according to the Floquet's theorem [1, 2, 23], can be expressed as:

$$\left.E_z\right|_{TM_{01-like}} = \underbrace{E_P(r,z)}_{\substack{\text{periodic function}\\\text{with period }D}} \cos(\omega_{RF} t - k_z^* z) e^{-\alpha z} \quad . \qquad (19)$$

In Eq. (19) the propagating constant does not have an analytical expression as in the case of constant cross-section waveguides and the dispersion curve for this type of structures is given in Fig. 11(b) and shows that, at a given frequency, the phase velocity can be equal to (or even slower than) *c*. In Eq. (19) we have also included the attenuation constant $\alpha$ *[1/m]* of the accelerating field due to the rf losses in the metallic walls. Typical values of α are in the range 0.2-0.3 1/m.

In a TW structure, the rf power enters into the cavity through an input coupler and flows through the cavity in the same beam direction. An output coupler, at the end of the structure, connected to a matched load, absorbs the residual power not transferred to the beam or dissipated in the cavity wall, avoiding reflections, as sketched in Fig. 11(c). If there is no beam, the input power simply dissipates on the cavity walls and the remainder is finally dissipated into the power load. In the presence of a beam current a fraction of this power is, indeed, transferred to the beam itself. TW structures allow acceleration over large distances (few meters, hundreds of cells) with just an input coupler and a relatively simple geometry.

For example, the SLAC electron Linac [24] is composed by more than one hundred 3 m long structures operating at 2.856 GHz while the SwissFEL linac by 96 structures, 2 m long operating at 5.712 GHz [25-27].

Similarly to what has been done for SW cavities, it is possible to define some figures of merit for TW structures as well. A complete description of these quantities is given, as example, in [1,2,19,28]. Here we want just to mention three quantities that are generally considered in the definition of the TW parameters. The first one is the attenuation constant $\alpha$, already mentioned, and defined as:

$$\alpha = \frac{p_{diss}}{2P_F} = -\frac{dP_F/dz}{2P_F} \quad , \tag{20}$$

where $p_{diss}$ is the dissipated power per unit length and $P_F$ is the power flowing into the structure at a given section. The second one is the shunt impedance per unit length $r$ defined as:

$$r = \frac{E_{acc}^2}{p_{diss}} \quad \left[\frac{\Omega}{m}\right] \quad , \tag{21}$$

where $E_{acc}$ is the average accelerating field. The higher the value of $r$, the higher the available accelerating field for a given rf power. Typical values for a 3 GHz structure are ~60 MΩ/m. The last quantity is the filling time $\tau_F$ defined as the time necessary to the rf wavefront to propagate from the input coupler to the end of a section of length $L$, and given by:

$$\tau_F = \frac{L}{v_g} \quad [s] \quad . \tag{22}$$

Typical values of the group velocity $v_g$ are 1-2% the speed of light and, as a consequence, the filling times are of the order of few hundreds on ns up to 1 μs. After one filling time the structure is completely filled of e.m. energy and the beam can be injected and efficiently accelerated. It can be demonstrated that the average accelerating field in a simple TW structure, can be expressed as $E_{acc} = \sqrt{2\alpha r P_{IN}} e^{-\alpha z}$ being $P_{IN}$ the input power.

In Appendix I it is also shown how the SW field of a multi-cell structure can be written as the sum of two counter propagating travelling waves.

## 8  Linac technology

An extensive dissertation on linac technology is given in [19]. Here we limit to point out that the linac structures can be realized with normal conducting or superconducting technology depending on the required linac performances in term of average accelerating field, type of particles, rf pulse length (i.e. how many bunches we can contemporary accelerate), duty cycle (i.e. pulsed operation or continuous wave operation) and average beam current.

The materials used for linac structure fabrication are: oxygen-free high conductivity (OFHC) copper for normal conducting structures (both SW and TW) and niobium for superconducting cavities (SW only). OFHC copper is the most common material used for NC structures because it has a very good electrical (and thermal) conductivity, it has a low Secondary Emission Yield (SEY), that allows to minimize multipacting phenomena [29] during structure power up, it shows good performance at high accelerating fields, it is easy to machine with very good roughness (up to the level of a few nm), it can be brazed or welded.

The most common material for the fabrication of SC cavities is niobium for several reasons [31-37]: it has a relatively high transition temperature ($T_c$=9.25 K), very low surface resistance when exposed to rf fields, it has a relatively high critical magnetic field, $H_c$=170-180 mT, it is chemically inert, can be machined and deep drawn, it is available either as bulk or sheet in any size, fabricated by forging and rolling, it can also be used as a coating (e.g. by sputtering) on NC materials like Cu; it has good thermal stability and has a relatively low cost.

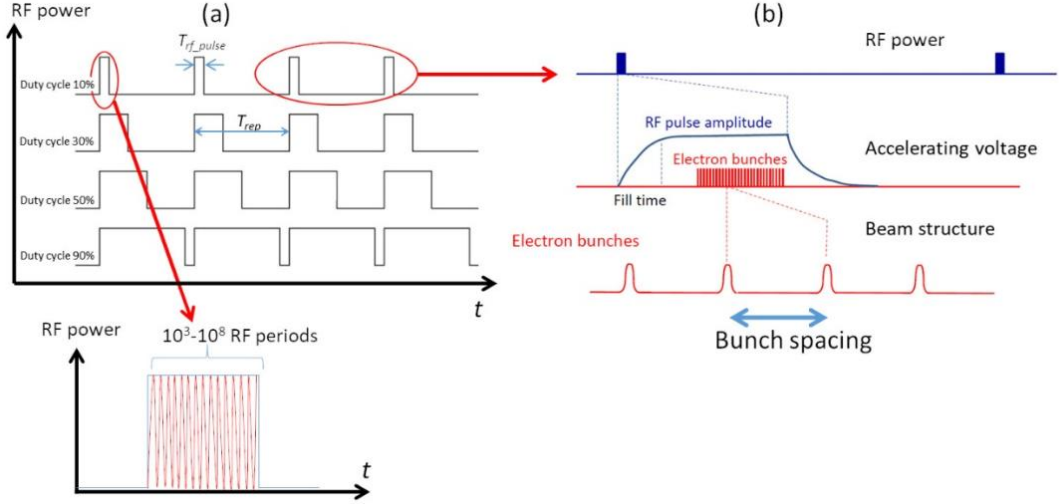

**Fig. 12:** (a) Sketch of the input rf power into accelerating structures and (b) related beam structure.

## 9 Beam structure and rf structure

Rf structures are fed, in general, by pulses with a certain repetition rate and duty cycle (*DC*), which is defined as the ratio between the pulse width ($T_{rf\_pulse}$) and the period ($T_{rep}$) ($DC=T_{rf\_pulse}/T_{rep}$). Each pulse includes from thousands up to several million rf periods as schematically represented in Fig. 12(a). The "beam structure" in a linac is correlated to the "rf structure" since a fraction of the rf pulse is used for beam acceleration (as represented in Fig. 12(b)). SC structures allow operation at very high DC (>1%) up to a continuous wave (CW) operation (DC=100%), because of the extremely low dissipated power, with relatively high accelerating field (>20 MV/m). This means that a continuous (bunched) beam can be accelerated. NC structures can operate in pulsed mode at lower DC (<0.1 %), because of the higher dissipated power with, in principle, larger peak accelerating field (>30 MV/m). In low DC linacs, depending on the application, from one up to few hundreds bunches can be, in general, accelerated.

## 10 Longitudinal beam dynamics of accelerated particles

In this second part of the paper the basic principles of the longitudinal and transverse beam dynamics will be illustrated. A complete dissertation is given as example in [1,2].

### 10.1 Synchronous phase

Let us consider a SW linac structure made by accelerating gaps (like a DTL) or cavities as shown in Fig. 13(a). In each gap the accelerating field oscillates in time and the accelerating voltage ($V_{acc}$) sketched in Fig. 13(b) can be expressed as:

$$V_{acc} = \hat{V}_{acc} \cos(\omega_{RF} t) \quad . \tag{23}$$

We can assume that the "perfect" synchronism condition is fulfilled for an ideal particle that crosses each gap with a phase $\phi_s$ with respect to the accelerating voltage. By definition this phase is called synchronous phase and the particle, synchronous particle. Then, the synchronous particle enters in each gap with a phase $\phi_s$ ($\phi_s = \omega_{RF} t_s$) with respect to the rf voltage, has an energy gain (and a consequent change in velocity) that allows to enter in the subsequent gap with the same phase $\phi_s$ with respect to the accelerating voltage and so on. For this particle the energy gain in each gap is simply given by:

$$\Delta E = q \underbrace{\hat{V}_{acc} \cos(\phi_s)}_{V_{acc\_s}} = q V_{acc\_s} \qquad (24)$$

Looking at the plot of Fig. 13 one case easily understand that both $\phi_s$ and $-\phi_s$ are synchronous phases. Let us consider the first synchronous phase $\phi_s$ (on the positive slope of the rf voltage). If we consider another "non-synchronous" particle "near" to the synchronous one, that arrives later in the gap ($t_1 > t_s$, $\phi_1 > \phi_s$), this particle will experience a higher voltage (i.e. gaining a slightly larger amount of energy) and thus will have an higher velocity than the synchronous one. Its time of flight to next gap will be reduced, partially compensating its initial delay with respect to the synchronous particle. Similarly, if we consider a particle that enters the gap before the synchronous one ($t_1 < t_s$, $\phi_1 < \phi_s$), it will experience a smaller accelerating voltage, gaining a smaller amount of energy and its time of flight to next gap will increase, compensating the initial advantage with respect to the synchronous particle. On the contrary, if we consider the synchronous particle at phase $-\phi_s$ and other particles "near" to it that arrive later or before in the gap, they will receive an energy gain that will increase further their distance with respect to the synchronous one. In conclusion, the synchronous phase on the positive slope of the rf voltage provides a longitudinal focusing of the beam allowing to have a stable beam acceleration. This mechanism is called "phase stability principle". On the contrary, the synchronous phase on the negative slope of the rf voltage is an unstable position.

Since it relies on particle velocity variations, longitudinal focusing does not work for fully relativistic beams (electrons). In this case acceleration "on crest" is more convenient.

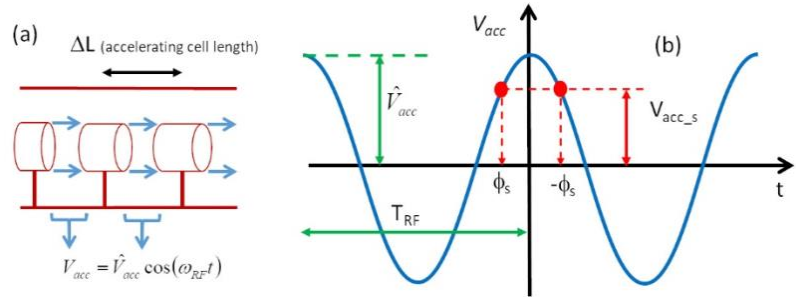

**Fig 13:** (a) Sketch of accelerating gaps; (b) accelerating voltage in each gap.

## 10.2 Energy-phase equations

The previous intuitive approach can find a more rigorous mathematical formalism. To this purpose, the following variables, $\varphi$ and $w$, are used to describe a generic particle longitudinal position with respect to the synchronous one. They are defined as follows:

$$\begin{cases} \varphi = \phi - \phi_s = \omega_{RF}(t - t_s) \\ w = W - W_s \end{cases}, \qquad (25)$$

where $t$ ($\phi$) is the arrival time (phase) of a generic particle at a certain gap and $W$ is the kinetic energy of the same particle at a certain position along the linac. $t_s$, $\phi_s$ and $W_s$ are the same quantities referred to the synchronous particle.

The energy gain per "accelerating cell" (considering as "accelerating cell" one gap plus the two half drift tubes in case of a DTL) of a generic particle and of the synchronous one are respectively:

$$\begin{cases} \Delta W_s = q\hat{V}_{acc} \cos\phi_s \\ \Delta W = q\hat{V}_{acc} \cos\phi = q\hat{V}_{acc} \cos(\phi_s + \varphi) \end{cases}. \tag{26}$$

The delta energy gain is then given by:

$$\Delta w = \Delta W - \Delta W_s = q\hat{V}_{acc}[\cos(\phi_s + \varphi) - \cos\phi_s] . \tag{27}$$

Dividing by the accelerating cell length $\Delta L$ and assuming that $\hat{E}_{acc} = \hat{V}_{acc}/\Delta L$ is the average accelerating field over the cell (i.e. average accelerating gradient) we obtain:

$$\frac{\Delta w}{\Delta L} = q\hat{E}_{acc}[\cos(\phi_s + \varphi) - \cos\phi_s] \Rightarrow \frac{dw}{dz} = q\hat{E}_{acc}[\cos(\phi_s + \varphi) - \cos\phi_s], \tag{28}$$

where we have approximated $\Delta w/\Delta L \approx dw/dz$.

On the other hand, the phase variation per cell of a generic particle and of a synchronous particle are:

$$\begin{cases} \Delta\phi_s = \omega_{RF}\Delta t_s \\ \Delta\phi = \omega_{RF}\Delta t \end{cases}, \tag{29}$$

where $\Delta t$ is basically the time of flight between two accelerating cells. Considering the variation of $\varphi$ between two accelerating gaps, $\Delta\varphi = \omega_{RF}(\Delta t - \Delta t_s)$, and dividing by $\Delta L$ we have that ([1]):

$$\frac{\Delta\varphi}{\Delta L} = \omega_{RF}\left(\frac{\Delta t}{\Delta L} - \frac{\Delta t_s}{\Delta L}\right) = \omega_{RF}\left(\frac{1}{v} - \frac{1}{v_s}\right) \underset{MAT}{\cong} -\frac{\omega_{RF}}{cE_0\beta_s^3\gamma_s^3} w \underset{\frac{\Delta\varphi}{\Delta L} \cong \frac{d\varphi}{dz}}{\Rightarrow} \frac{d\varphi}{dz} = -\frac{\omega_{RF}}{cE_0\beta_s^3\gamma_s^3} w . \tag{30}$$

The system of coupled (non-linear) differential equations represented by Eqs. (28) and (30), and reported in the following, describes the motion of a non-synchronous particle in the longitudinal plane with respect to the synchronous one:

---

[1] In the MAT passage we have considered the following approximations:

$$\omega_{RF}\left(\frac{1}{v} - \frac{1}{v_s}\right) = \omega_{RF}\left(\frac{v_s - v}{vv_s}\right) \underset{\substack{vv_s \cong v_s^2 \\ v - v_s \cong \Delta v}}{\cong} -\frac{\omega_{RF}}{v_s^2}\Delta v = -\frac{\omega_{RF}}{c}\frac{\Delta\beta}{\beta_s^2} \; ; \; \beta = \sqrt{1 - 1/\gamma^2} \Rightarrow \beta d\beta = d\gamma/\gamma^3 \Rightarrow -\frac{\omega_{RF}}{c}\frac{\Delta\beta}{\beta_s^2} \cong -\frac{\omega_{RF}}{c}\frac{\Delta\gamma}{\beta_s^3\gamma_s^3} = -\frac{\omega_{RF}}{c}\frac{\frac{w}{\Delta E}}{E_0\beta_s^3\gamma_s^3}$$

$$\begin{cases} \dfrac{dw}{dz} = q\hat{E}_{acc}[\cos(\phi_s + \varphi) - \cos\phi_s] \\ \dfrac{d\varphi}{dz} = -\dfrac{\omega_{RF}}{cE_0\beta_s^3\gamma_s^3}w \end{cases} \tag{31}$$

## 10.3 Small amplitude energy-phase oscillations

Deriving the second equation of (31) with respect to z and assuming an adiabatic acceleration process (i.e. $[d(\omega_{RF}/cE_0\beta_s^3\gamma_s^3)/dz]w \ll (\omega_{RF}/cE_0\beta_s^3\gamma_s^3)dw/dz$ ) we obtain:

$$\frac{d^2\varphi}{dz^2} = -\frac{\omega_{RF}}{cE_0\beta_s^3\gamma_s^3}\frac{dw}{dz} \tag{32}$$

Assuming in the first equation of (31) small oscillations around the synchronous particle that allow to approximate $\cos(\phi_s + \varphi) - \cos\phi_s \cong \varphi\sin\phi_s$ and substituting this equation in (32) we finally obtain:

$$\frac{d^2\varphi}{dz^2} + \underbrace{q\frac{\omega_{RF}\hat{E}_{acc}\sin(-\phi_s)}{cE_0\beta_s^3\gamma_s^3}}_{\Omega_s^2}\varphi = 0 \tag{33}$$

This is the equation of a harmonic oscillator with angular spatial frequency $\Omega_s$. The conditions to have stable longitudinal oscillations and acceleration at the same time are then:

$$\left.\begin{array}{l}\Omega_s^2 > 0 \Rightarrow \sin(-\phi_s) > 0 \\ V_{acc} > 0 \Rightarrow \cos\phi_s > 0\end{array}\right\} \Rightarrow -\frac{\pi}{2} < \phi_s < 0 \tag{34}$$

Equation (33) and the condition (34) summarizes what we have intuitively described with the "phase stability principle": if we accelerate on the rising part of the positive rf wave we have a longitudinal force keeping the beam "focused" around the synchronous phase and oscillating during acceleration with spatial angular frequency $\Omega_s$. The angular frequency can be simply obtained substituting in Eq. (33) z with $\beta_s ct$ (and then $dz=\beta_s cdt$) and it is related to the angular spatial frequency by: $\Omega_T = \Omega_s\beta_s c$.

Each particle during acceleration describes in the longitudinal plane, an ellipse around the synchronous particle, as schematically represented in Fig. 14.

The maximum energy deviation is reached at $\varphi=0$ while the maximum phase excursion corresponds to $w=0$.

All particles in the bunch occupy an area in the longitudinal phase space called longitudinal emittance and the projections of the bunch in the energy and phase planes give the energy spread and the bunch length as schematically represented in Fig. 14.

From Eq. (33) it is also evident that the angular frequency scale with $1/\gamma^{3/2}$ that means that for ultra-relativistic electrons shrinks to zero and the beam is "frozen" and does not oscillate in the longitudinal plane.

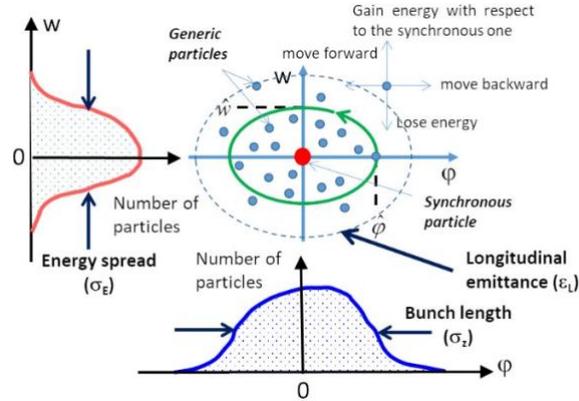

**Fig 14:** oscillating particles in the longitudinal phase space around the synchronous one.

## 11  Longitudinal dynamics of low energy electrons

From previous formulae it is clear that there is no motion in the longitudinal phase space for ultra-relativistic particles with $\gamma \gg 1$. This is the case of electrons, whose velocity is always close to the speed of light even at low energies. For electrons, accelerating structures are designed to provide an accelerating field synchronous with particles moving at $v=c$, as in the case of TW structures with phase velocity equal to c.

It is interesting to analyze what happens if we inject an electron beam generated in an electron gun (at low energy) directly in a TW structure (with $v_{ph}=c$) and the conditions that allow to capture the beam ([2]). The sketch is given in Fig. 15. Particles enter in the TW structure with velocity v<c and, initially, they are not synchronous with the accelerating field. In this part of the accelerator there is a so-called "phase slippage". After a certain distance they can reach enough energy (and velocity) to become synchronous with the accelerating wave (Fig. 15(a)). This means that they are captured by the accelerator and, from this point on, they are permanently accelerated. If this does not happen (i.e. the energy increase is not enough to reach the velocity of the wave) they are lost.

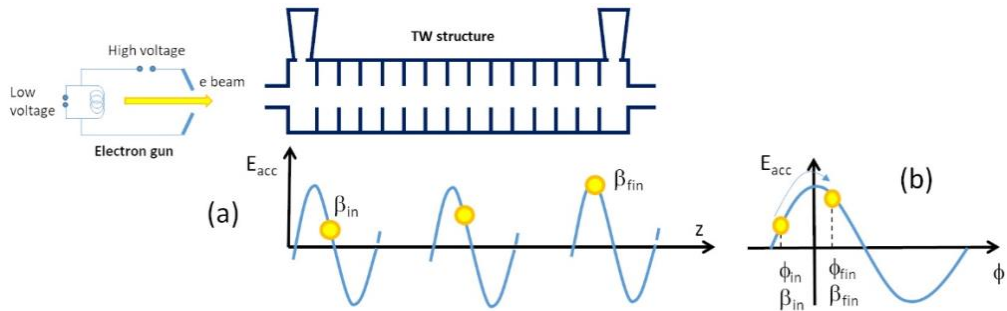

**Fig. 15:** basic scheme of a low energy electron gun coupled with a TW accelerating structure: capture process (a) and phase slippage (b).

### 11.1  Phase slippage

We will now describe with a more precise mathematical approach this phenomenon. The accelerating field of a TW structure can be expressed by:

---

[2] This is equivalent to consider instead of a TW structure a SW designed to accelerate ultra-relativistic particles at v=c

$$E_{acc} = \hat{E}_{acc} \cos\underbrace{(\omega_{RF} t - kz)}_{\phi(z,t)}, \tag{35}$$

where $\phi$ represents the phase of the particle with respect to the accelerating wave.

The equation of motion of a particle with a position z at time t accelerated by the TW is then, from the Lorentz force, given by:

$$\frac{d}{dt}(mv) = q\hat{E}_{acc}\cos\phi(z,t) \Rightarrow m_0 c \frac{d}{dt}(\gamma\beta) = m_0 c \gamma^3 \frac{d\beta}{dt} = q\hat{E}_{acc}\cos\phi. \tag{36}$$

Integrating both terms of Eq. (36) between an initial and a final state [1,38] we can find which is the relation between $\beta$ and $\phi$ from an initial condition to a final one:

$$\sin\phi_{fin} = \sin\phi_{in} + \frac{2\pi E_0}{\lambda_{RF} q\hat{E}_{acc}}\left(\sqrt{\frac{1-\beta_{in}}{1+\beta_{in}}} - \sqrt{\frac{1-\beta_{fin}}{1+\beta_{fin}}}\right). \tag{37}$$

Supposing that the particle reaches, asymptotically, the value $\beta_{fin}=1$ we have:

$$\sin\phi_{fin} = \sin\phi_{in} + \frac{2\pi m_0 c^2}{\lambda_{RF} q\hat{E}_{acc}}\sqrt{\frac{1-\beta_{in}}{1+\beta_{in}}}. \tag{38}$$

Equation (38) (or the more general Eq. (37)) gives several information on the physics of the low energy acceleration process. First of all, in order to have a solution for the final phase $\phi_{fin}$, the second term of the Eq. (38) should be in the interval [-1,1]. This means that, for a given initial injection phase, we always have that $\sin\phi_{fin} > \sin\phi_{in}$ that means $\phi_{fin} > \phi_{in}$. This is the mathematical description of the phase slippage phenomenon as represented in Fig. 15 (b). Another important consequence of Eq. (38) is that, for a given accelerating field and rf frequency, there are only some possible injection phases for which we can capture the beam. Conversely, for a given injection energy ($\beta_{in}$) and phase $\phi_{in}$ we can find which is the accelerating peak field ($\hat{E}_{acc}$) that is necessary, to have a relativistic beam at phase $\phi_{fin}$ (i.e. that is necessary to capture the beam at phase $\phi_{fin}$).

## 11.2  Bunch compression

Equation (38) is useful to describe the bunch compression during the capture process. As the injected beam moves up to the crest, in fact, it experiences also a longitudinal bunching, which is caused by velocity modulation (hence the name "velocity bunching"). This is evident plotting the final phases as a function of the injection phases for different accelerating field $\hat{E}_{acc}$ as reported, for example, in Fig. 16(a). Differentiating Eq. (37) it is straightforward to derive the first order compression factor:

$$\Delta\phi_{fin} = \Delta\phi_{in} \frac{\cos\phi_{in}}{\cos\phi_{fin}}. \tag{39}$$

This mechanism can be used to compress the electron bunches in the first stages of acceleration [39].

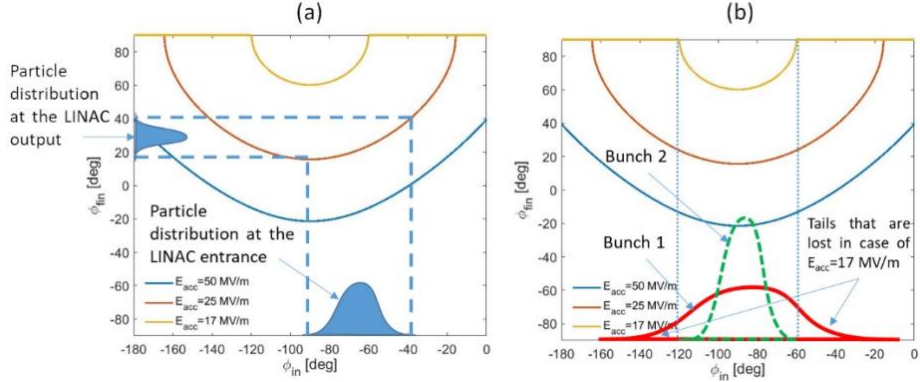

**Fig. 16:** (a) Final phase as a function of the injection phase for different accelerating field, assuming $f_{RF}$=3 GHz with two sketched bunches before and after the capture process; (b) Scheme of two bunches with different length captured by a TW structure coupled to a thermionic gun (the tails of the longer one are not captured by the low accelerating field).

### 11.3 Capture efficiency and buncher

For a given $E_{acc}$ we can easily calculate the range of the injection phases $\phi_{in}$ actually accepted in the capture process (i.e. particles whose injection phases are within this range can be captured while the other are lost). Fig. 16(b) illustrate this concept. Assuming an accelerating voltage of $E_{acc}$=17 MV/m (and $f_{RF}$=3 GHz), only the electrons that enter into the TW structure with a phase between -120 and -60 deg are captured, the other are lost. To reduce the number of lost particles there are, in principle two possibilities. The first one is to increase the accelerating voltage, thus increasing the range of phases. The other is to pre-shape the beam in order to increase the number of electrons that occupy the right phase interval. While the first solution requires more rf power to increase the accelerating field, this second approach can be pursued with a simple scheme as reported in Fig. 17, where a typical injector scheme is reported. The scheme foresees the use of a buncher. It is a SW cavity aimed at pre-forming the particle bunch gathering particles continuously emitted by a source by modulating the energy (and therefore the velocity) of the continuous emitted beam, using the longitudinal E-field of the SW cavity itself. After a certain drift space, the velocity modulation is "converted" in a density charge modulation. The density modulation depletes the regions corresponding to injection phase values incompatible with the capture process. The TW accelerating structure (capture section) is placed at an optimal distance from the buncher, to capture a large fraction of the charge and accelerate it up to relativistic energies. The amount of charge lost using this scheme is drastically reduced, while the capture section provides also further beam bunching.

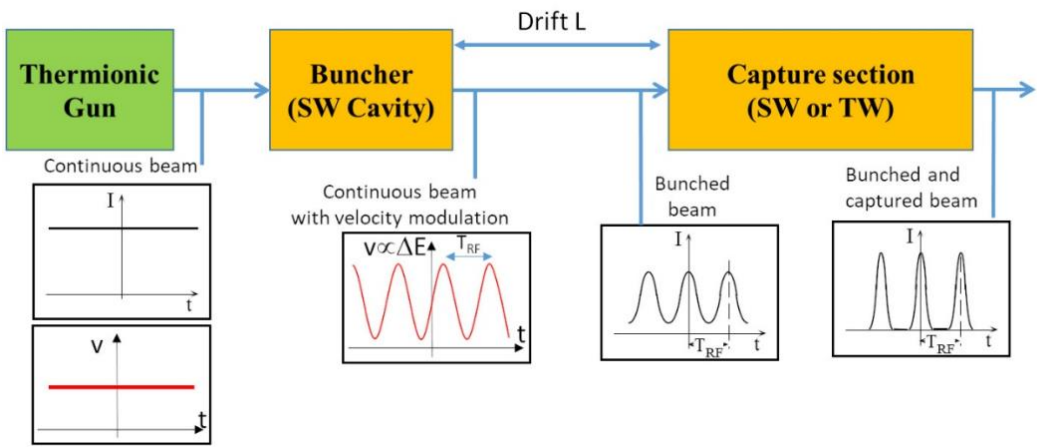

**Fig. 17:** Scheme of a typical electron injector using a buncher.

## 12 Transverse beam dynamics of accelerated particles

We will now describe more in detail the transverse motion of particles that are accelerated in a linac. The transverse beam dynamics is determined by the effect of the rf fields, by the magnetic elements (quadrupoles or solenoids) and by the collective effects such as space charge forces and wakefields.

### 12.1 Rf transverse forces: defocusing term

The rf fields that accelerate particles act also on the transverse beam dynamics, because of the off axis transverse components of the electric and magnetic field, as schematically represented in Fig. 18(a), where the electric field lines in a generic accelerating gap are reported. More precisely, considering the Maxwell's equations in vacuum and assuming an accelerating SW field of the type reported in Eq. (10), we can calculate the transverse force, as reported in the following:

$$\begin{cases} \nabla \vec{E} = 0 \\ \nabla \times \vec{B} = \dfrac{1}{c^2} \vec{E} \end{cases} \underset{\text{cylindrical coordinates}}{\Rightarrow} \begin{cases} E_r = -\dfrac{r}{2}\dfrac{\partial E_z}{\partial z} \\ B_\theta = \dfrac{r}{2c^2}\dfrac{\partial E_z}{\partial t} \end{cases} \underset{\text{Lorentz Force}}{\Rightarrow} \dfrac{F_r}{q} = E_r - vB_\theta = \begin{cases} \underbrace{-\dfrac{r}{2}\dfrac{\partial E_{RF}(z)}{\partial z}\cos\left(\omega_{RF}\dfrac{z}{\beta c}+\phi\right)}_{\text{E contribution}} + \\ \underbrace{\dfrac{r}{2}\omega_{RF}\dfrac{\beta}{c}E_{RF}(z)\sin\left(\omega_{RF}\dfrac{z}{\beta c}+\phi\right)}_{\text{B contribution}} \end{cases}. \tag{40}$$

As an example, the transverse forces as a function of the longitudinal coordinate are reported in Fig. 18(b) for an accelerating gap of L= 3 cm, working at $f_{RF}$=350 MHz, for a β=0.1 and for two different injection phases. From the plot it is evident that the transverse forces are equal to zero in the center of the gap (except $F_r|_B$ in the case φ≠0), while they have an opposite sign at the entrance and at the exit of the gap itself, as also visible in the schematic picture of Fig. 18(a). It is also evident that the electric field contribution is dominant and that, if the injection phase is negative (as required for longitudinal focusing) the two in/out contributions do not compensate, resulting in an integrated defocusing force.

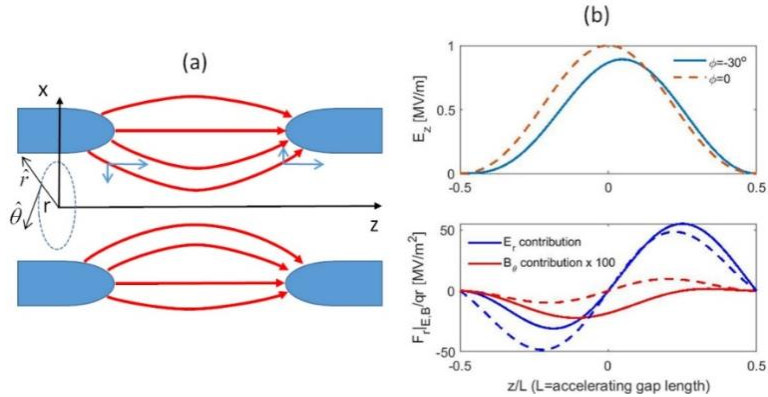

**Fig. 18:** (a) Electric field lines in a generic accelerating gap; (b) transverse forces as a function of the longitudinal coordinate in traversing the accelerating gap.

More precisely, from the previous formulae it is possible to calculate the transverse momentum increase due to the rf transverse forces. Assuming that the velocity and particle position do not change across the gap, we obtain to the first order:

$$\Delta p_r = \int_{-L/2}^{+L/2} F_r \dfrac{dz}{\beta c} = -\dfrac{\pi q \hat{E}_{acc} L \sin\phi}{c \gamma^2 \beta^2 \lambda_{RF}} r. \tag{41}$$

The formula highlights the defocusing nature of such term (since $sin\phi<0$), that scales as $1/\gamma^2$. As a consequence, it disappears at relativistic regime (i.e. for electrons[3]). For a correct evaluation of the defocusing effect in the non-relativistic regime we have also to take into account the velocity change across the accelerating gap, the transverse beam dimensions changes across the gap (with a general reduction of the transverse beam dimensions due to the focusing in the first part). Both contributions give a reduction of the defocusing force but the resulting one is still defocusing.

## 12.2  Rf focusing in electron linacs

We have pointed out that the rf defocusing term is negligible in electron linacs. It is important to mention that, for this type of linacs, there is a second order effect due to the non-synchronous harmonics of the accelerating field that give a net focusing contribution [40]. These harmonics generate a ponderomotive force i.e., a force in an inhomogeneous oscillating electromagnetic field. As discussed in [40] it is possible to demonstrate that we have an average focusing force given by this expression:

$$\bar{F}_r = -rq\frac{\hat{E}_{acc}^2}{8\gamma\, m_0 c^2/e}\eta(\phi), \qquad (42)$$

where the term $\eta(\phi)$ is a factor that depends on the harmonic content of the accelerating field and is of the order of 0.1 [40]. This harmonic content is larger in SW cavities because of the presence of the wave that propagate in the opposite direction with respect to the beam (as pointed out in Appendix I). With accelerating gradients of few tens of MV/m it is quite easy to verify that this transverse gradient can easily reach the level of MV/m$^2$.

## 12.3  Collective effects: space charge forces

Collective effects are phenomena related to the number of particles in a bunch, and, in linacs, they can play a crucial role in the longitudinal and transverse dynamics of intense beams. They are typically related to space charge effects and wakefield and here we will focus only on the former. This effect is generated by the Coulomb repulsion between particles. If we consider a uniform and infinite cylinder of charge with radius $R_b$ moving along the longitudinal axis z with an average current $I$, it is quite easy to verify that the force experienced by a generic particle inside the cylinder at a radius $r_q$ is given by:

$$\vec{F}_{SC} = q\frac{I}{2\pi\varepsilon_0 R_b^2 \beta c\gamma^2}r_q\hat{r}. \qquad (43)$$

This simple example gives us a couple of interesting information: the effect of space charge is of particular concern for low energy particles and high currents, because the space charge forces scale as $1/\beta\gamma^2$ and linearly with the current I. In this particular example the force is linear with the displacement $r_q$, but in general space charge forces are non-linear and depend on the beam particle distribution.

## 12.4  Magnetic focusing, transverse oscillations and $\beta$-function

In the previous paragraphs we pointed out that the rf and space charge forces (or the natural divergence of the beam given by a generic source) give a defocusing effect and need to be compensated, in order to take under control, the transverse beam dynamics in a linac. For this reason, quadrupoles are used to focus the beam and, at low energies, also solenoids. We will consider, in particular, the first type of elements. As done in all accelerator machines with strong focusing magnets, quadrupoles are used in an alternating configuration, since they are focusing in one plane and defocusing into the other. In a linac they are interleaved by either accelerating gaps or accelerating structures. The type of magnetic configuration and the magnets distance depend on the type of particles, energy, beam intensity and beam dimensions requirements. Due to the alternating quadrupole focusing system, each particle (as in synchrotrons) performs transverse oscillations and the equation of motion in the transverse plane is of the type:

$$\frac{d^2x}{ds^2} + \underbrace{\left[\overbrace{\kappa^2(s)}^{\text{magnetic focusing}} - \overbrace{k_{RF}^2(s)}^{\text{RF defocusing}}\right]}_{K^2(s)} x - F_{SC} = 0 \quad . \tag{44}$$

---

[3] In the case of electrons, moreover, we already pointed out that, in general, $\phi=0$ for maximum acceleration and this also cancel the defocusing effect.

The *K(s)* term takes into account the magnetic configuration and the rf defocusing term that exhibits a linear behavior with the particle displacement, while the $F_{SC}$ term is the non-linear space charge term.

If we neglect the space charge forces and we assume $K^2>0$, the solution of the single particle trajectory described by the Eq. (44) is a pseudo-sinusoid described by the equation ([4]):

$$x(s) = \sqrt{\varepsilon_o \beta(s)} \cos\left[\int_{s_0}^{s} \frac{ds}{\beta(s)} + \phi_0\right] , \tag{45}$$

where the characteristic *β*-function depend on the magnetic and rf configuration along the linac and the constants $\varepsilon_0$ and $\phi_0$ depend on the initial conditions of the particle at the entrance of the linac (i.e. position and angle). Fig. 19(a) shows, as an example, different particle trajectories corresponding to different initial conditions. It is quite easy to verify that, because of Eq. (45), all particles oscillations are contained within an envelope that scales with the square root of the *β*-function, sketched, as an example in Fig 19(a) and also the final transverse beam dimensions ($\sigma_{x,y}(s)$) vary along the linac within the same envelope.

Because of the regular magnetic and accelerating structure configuration, the *β*-function is "locally" periodic, and it is useful to refer to the focusing period ($L_p$) that is the length after which the focusing structure is repeated (usually equal to $N\beta\lambda$). The transverse phase advance per period is defined as:

$$\sigma = \int_{L_p} \frac{ds}{\beta(s)} \approx \frac{L_p}{\langle\beta\rangle} \tag{46}$$

And, for transverse oscillation stability, should be in the range $0<\sigma<\pi$ [1]. The quantity $\sigma/L_p$ is called phase advance per unit length.

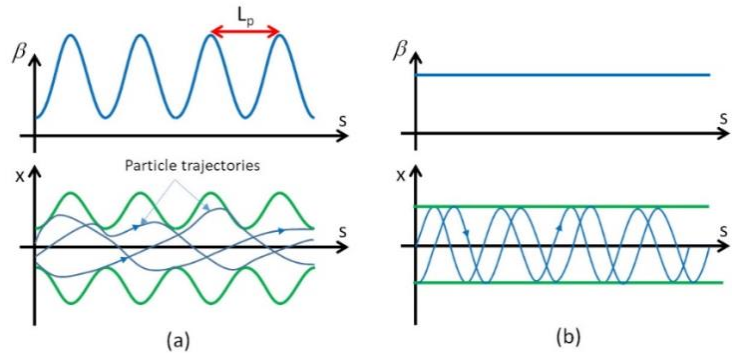

**Fig. 19:** (a) Sketch of the *β*-function and transverse particle trajectories along the linac; (b) *β*-function and transverse particle trajectories in the case of "smooth approximation".

---

[4] Unfortunately, the Twiss function β here has the same notation than the relativistic factor β. They are, obviously, two completely different quantities.

## 12.5 Smooth approximation of the transverse oscillations

In case of "smooth approximation" of the linac, we consider an average effect of the quadrupoles and rf and, as a consequence, a constant $\beta$-function. Assuming a focusing structure of the type sketched in Fig 20, which alternates focusing quadrupoles with accelerating gaps (also called FODO lattice) we obtain a simple harmonic motion along s coordinate of the type:

$$x(s) = \sqrt{\varepsilon_o} \sqrt{1/K_0} \cos(K_0 s + \phi_0) \quad . \tag{47}$$

In particular for this simple case, we obtain that the phase advance per unit length $K_0$ is equal to [1]:

$$K_0 = \sqrt{\underbrace{\left(\frac{qGl}{2m_0 c\gamma\beta}\right)^2}_{\text{magnetic focusing term}} - \underbrace{\frac{\pi q \hat{E}_{acc} \sin(-\phi)}{m_0 c^2 \lambda_{RF}(\gamma\beta)^3}}_{\text{RF defocusing term}}} \quad , \tag{48}$$

where G [T/m] is the quadrupole gradient and $l$ [m] is the quadrupole length, and all other quantities have been already defined in the previous sections ([5]). In the previous formula it is possible to recognize the two contributions of the magnetic focusing and of the rf defocusing (with opposite sign with respect to the previous one). This simplified approach allows to make several considerations. First of all, the rf defocusing term scales with $f_{RF}$ ($=1/\lambda_{RF}$), and this sets a higher limit to the working frequency of the cavities (i.e. at lower particle energies it is better to operate at lower frequency). Moreover, as already pointed out, the rf defocusing term plays a crucial role at low energy, since the defocusing term scale as $1/(\gamma\beta)^3$.

If we consider also the space charge contribution and the simple case of an ellipsoidal beam of charge Q (that gives linear space charge forces) we obtain for $K_0$ the following expression [1]:

$$K_0 = \sqrt{\underbrace{\left(\frac{qGl}{2m_0 c\gamma\beta}\right)^2}_{\text{magnetic focusing term}} - \underbrace{\frac{\pi q \hat{E}_{acc} \sin(-\phi)}{m_0 c^2 \lambda_{RF}(\gamma\beta)^3}}_{\text{RF defocusing term}} - \underbrace{\frac{3Z_0 q I \lambda_{RF}(1-f)}{8\pi m_0 c^2 \beta^2 \gamma^3 r_x r_y r_z}}_{\text{space charge defocusing term}}} \quad , \tag{49}$$

where $I$ is the average beam current ($=Q/T_{RF}$), $r_{x,y,z}$ are the ellipsoid semi-axis, $f$ is a form factor ($0<f<1$) and $Z_0$ is the free space impedance (equal to 377 $\Omega$). For ultra-relativistic particles (e.g., electrons of several MeV) both the rf defocusing and the space charge terms disappear and the external focusing is only required to control the emittance and beam dimensions and to stabilize the beam against instabilities.

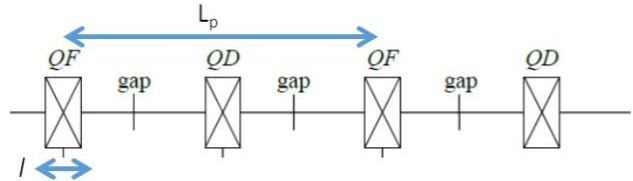

**Fig 20:** Focusing structure considered in the smooth approximation (FODO lattice).

---

[5] Please note that $\beta$ in this formula is the relativistic factor.

## 12.6 General considerations on linac optics design for protons and ions

According to what illustrated, the beam dynamics is completely dominated by space charge and rf defocusing forces. Focusing is usually provided by quadrupoles that are also integrated in the drift tubes of DTL structures. The phase advance per period ($\sigma$) should be, in general, in the range 30-80 deg [1,2]. This means that, at low energy, we need a strong focusing term (short quadrupole distance and high quadrupole gradient) to compensate for the rf defocusing, but the limited space in the drift tubes (proportional to $\beta\lambda$) limits this achievable integrated gradient and, as a consequence, the beam current. As $\beta$ increases, the distance between focusing elements can increase: $\beta\lambda$ in the DTL goes, as example, from ~70 mm at 3 MeV, $f_{RF}$=350 MHz to ~250 mm at 40 MeV and can be increased to 4-10 $\beta\lambda$ at higher energy (>40 MeV). As already pointed out, the overall linac is made of a sequence of structures, matched to the beam velocity, and where the length of the focusing period increases with energy. As $\beta$ increases, longitudinal phase error between cells of identical length becomes small, and we can have short sequences of identical cells (instead of cells all with different dimensions) with a reduction of the overall construction costs. From beam dynamics simulations it is possible to calculate the beam radius along the structures that allows to calculate the margin between beam radius and physical apertures.

## 12.7 General considerations on linac optics design for electrons

For electrons, space charge forces act only at low energy or high peak current. Below 10-20 MeV (injector) the beam dynamics optimization has to include emittance compensation schemes using, typically, solenoids. At higher energies, no space charge and no rf defocusing effects occur, but we have rf focusing due to the ponderomotive force in accelerating structures. In this part of the linac focusing periods up to several meters can occur. Nevertheless, the optics design has to take into account longitudinal and transverse wakefields (due to the higher frequencies used for acceleration) that can cause energy spread increase, head-tail oscillations and multi-bunch instabilities. The longitudinal bunch compression schemes based on magnets and chicanes have also to take into account, for short bunches, the interaction between the beam and the emitted synchrotron radiation in bending magnets (Coherent Synchrotron Radiation effects-CSR) [41,42]. All these effects are important especially in linacs for Free Electron Lasers (FEL) that require extremely good beam qualities, short bunches and high peak current.

## 13 Radio Frequency quadrupoles (RFQ)

At low proton (or ion) energies ($\beta$~0.01), space charge defocusing is strong and quadrupole focusing is not very effective (because of the low $\beta$). Moreover, cell lengths become small and conventional accelerating structures (DTL) are very inefficient. At these energies, it is used another type of structure, called Radio Frequency Quadrupole (RFQ), invented in the '60 by Kapchinskiy and Tepliakov [43]. These structures allow to simultaneously provide transverse focusing, acceleration and bunching of the beam. The sketch of the structure and the picture of a fabricated one are given in Fig. 21.

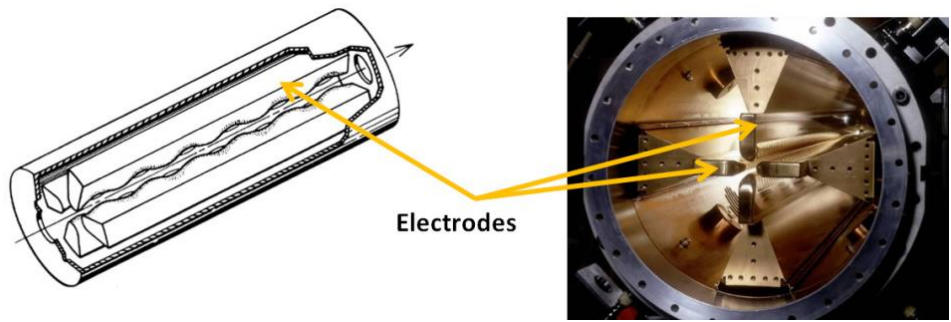

**Fig. 21:** Sketch of a RFQ structure and picture of a fabricated one.

The transverse focusing effect is due to the fact that the resonating mode of the cavity (between the four electrodes) is the quadrupole mode ($TE_{210}$) whose electric field lines are sketched in Fig. 22(a). The working principle of a quadrupole electric mode is similar to that of a magnetic quadrupole with no field in the center of the structure and a transverse electric field that increases linearly with beam off-axis and that is focusing in one plane and defocusing in the other. The rf alternating voltage on the electrodes produces an alternating focusing channel with a period rf.

The acceleration process is achieved by means of a longitudinal modulation of the vanes with period $\beta\lambda_{RF}$. This creates a longitudinal component of the electric field (as given in Fig. 22(b)) that accelerates the beam (the modulation corresponds exactly to a series of rf gaps).

The third effect, i.e. the bunching, is generated by changing the modulation period (distance between electric field maxima), since it is designed to change the phase of the beam with respect to the rf field during beam acceleration, while the amplitude of the modulation can be varied to change the accelerating gradient. The continuous beam enters the first cell and it is bunched around the -90 deg phase (bunching cells), progressively the beam is bunched and accelerated (adiabatic bunching channel) and, only in the last cells we have a switch on a pure acceleration process. The process is illustrated in Fig. 22(c).

The mathematical description of this process would require a dedicated course and is out of the scope of the present proceeding. Details can be found in [1,44].

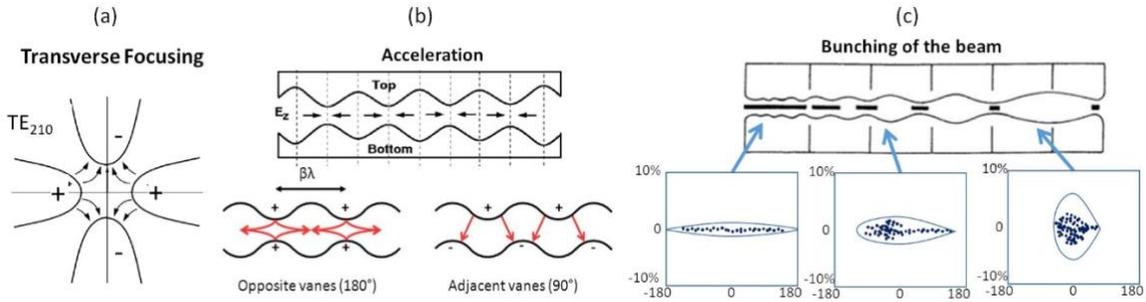

**Fig. 22:** (a) Quadrupole mode ($TE_{210}$) of the RFQ four vane geometry; (b) Longitudinal modulation of the electrodes with sketched electric field lines; (c) Bunching mechanism in RFQ.

## 14 Choice of the frequency of operation and accelerating structures

The choice of the frequency and type of accelerating structures depends on several factors such as: type of particles to be accelerated, average beam current and duty cycle, available space and compactness, cost.

Table I schematically illustrates how the accelerating structure parameters scale with frequency [19]. For NC structures *r* increases with frequency as $f^{1/2}$ and this pushes to adopt higher frequencies. Nevertheless, at high frequencies the beam-cavity interaction due to wakefields becomes more critical ($w_z \propto f^2$, $w_\perp \propto f^3$). On the other hand, for SC structures it can be demonstrated that the power losses increase with $f^2$ and, as a consequence, *r* scales with *1/f*, this pushes to adopt lower frequencies.

Higher frequencies allow to achieve, in general, higher accelerating gradients in NC structures but require higher mechanical precision in the realization of the structures. Moreover, at very high frequencies (>10-12 GHz) power sources are either commercially not available or very expensive. On the other hand, at low frequencies one needs more bulk material and machines have generally a larger footprint.

In DTL, for protons and ions, the accelerating cell dimensions (that scale as $\sim c\beta/f_{RF}$) become not practical at high frequency (as example at 3 GHz and beta=0.1 the accelerating cells are 10 mm) and also the insertion of magnetic elements in the drift tubes for transverse focusing is not possible. Also, the rf defocusing effects scale as *1/f* and they do not allow a stable acceleration of the beam.

In general, a given accelerating structure has a curve of efficiency (shunt impedance per unit length) with respect to the particle energy and, if it works at given β, the same structure, becomes not efficient at larger β and the transition to another type of structure is necessary.

As a general consideration normal conducting linacs require high peak power from power sources and also high average power in case of operation at relatively high DC (0.1-1%). For high duty cycle linacs (>1%) the use of superconducting structures is the only possible solution.

For all previous considerations, it is possible to make some general considerations schematically reported in Table II. In low duty cycle ($<10^{-3}$) electron linacs higher frequencies (up to X band) can be used for both SW than TW structures. Proton and ion linacs use low frequencies (40-500 MHz) up to β~0.5 with RFQ or DTL structures. At higher energies (β>0.5) SW multi cell structures at higher operating frequencies are used (from 500 MHz up to few GHz).

Superconducting multi cell cavities working on the π-mode, are used in high duty cycle linacs starting from β larger than 0.5 and, in case of protons or ions, combined with low frequency structures in the first stages of acceleration. A complete review of superconducting accelerating cavities is given in [45].

**Table I:** Scaling laws for cavity parameters with frequency

| Parameter | NC | SC |
|---|---|---|
| Surface resistance ($R_s$) | $\propto f^{1/2}$ | $\propto f^{2}$ |
| Quality factor ($Q$) | $\propto f^{-1/2}$ | $\propto f^{-2}$ |
| Shunt impedance per unit length ($r$) | $\propto f^{1/2}$ | $\propto f^{-1}$ |
| $r/Q$ | $\propto f$ | |
| Longitudinal wakefield ($w_{//}$) | $\propto f^{2}$ | |
| Transverse wakefield ($w_{\perp}$) | $\propto f^{3}$ | |

**Table II:** typical use of different accelerating structures (not exhaustive)

| Cavity Type | β Range | Frequency | Particles type |
|---|---|---|---|
| RFQ | 0.01– 0.1 | 40-500 MHz | Protons, Ions |
| DTL | 0.05 – 0.5 | 100-500 MHz | Protons, Ions |
| Multi cell (π or π/2 cavities) | 0.5 – 1 | 600 MHz-3 GHz | Protons, Electrons |
| SC multi cell π-mode | > 0.5-1 | 350 MHz-3 GHz | Protons, Electrons |
| TW | 1 | 3-12 GHz | Electrons |

## Acknowledgements

I would like to thank Luca Piersanti for the careful reading, corrections and helpful suggestions. Several pictures, schemes, images and plots have been taken from papers and presentations reported in the References: I would like to acknowledge all the authors.

## Appendix I: SW field as a sum of two counter-propagating TW waves.

In a multi cell SW structure working on the π-mode, the accelerating field can be expressed in a simplified form, as:

$$E_z = E_{RF}(z)\cos(\omega_{RF} t) = E_z = \underbrace{\hat{E}_{RF}\cos(k_z z)}_{E_{RF}(z)}\cos(\omega_{RF} t) \quad, \tag{A1}$$

where we have considered a simple expression for $E_{RF}(z)$. In order to have synchronism between the accelerating field and the particle of velocity $v$, and supposing that the velocity variation of the particle while traversing the structure is negligible, $k_z$ has to satisfy the following relation:

$$k_z = \frac{2\pi}{v T_{RF}} = \frac{2\pi}{\beta \lambda_{RF}} = \frac{\omega_{RF}}{v} \quad. \tag{A2}$$

The accelerating field seen by the particle is then given by ($t=z/v$):

$$E_z\Big|_{\substack{\text{seen} \\ \text{by} \\ \text{particle} \\ z=vt}} = \hat{E}_{RF}\cos(k_z z)\cos\left(\omega_{RF}\frac{z}{v}\right) = \hat{E}_{RF}\cos^2(k_z z) = \frac{\hat{E}_{RF}}{2} + \frac{\hat{E}_{RF}}{2}\cos(2k_z z) \quad. \tag{A3}$$

On the other hand, the SW can be written, with some math, as the sum of two TWs in the form:

$$E_z = \hat{E}_{RF}\cos(k_z z)\cos(\omega_{RF} t) = \frac{\hat{E}_{RF}}{2}\cos(\omega_{RF} t - k_z z) + \frac{\hat{E}_{RF}}{2}\cos(\omega_{RF} t + k_z z) \quad. \tag{A4}$$

Substituting in this last expression $t=z/v$ we obtain the same expression (A3). In conclusion the SW field can be "seen" as the superposition of two counter propagating TW waves. The co-propagating one is those that gives the net acceleration $E_z = \hat{E}_{RF}/2$, while the other one (back propagating) does not contribute to the acceleration but generates an oscillating term with no net acceleration effect.

## Appendix II: Large longitudinal oscillations and separatrix

To study the longitudinal dynamics at large oscillations, we have to consider the non-linear system of differential equations without approximations as given in Eq. (31). In the adiabatic acceleration case it is possible to obtain the following relation between w and φ [1]:

$$\frac{1}{2}\left(\frac{\omega_{RF}}{cE_0\beta_s^3\gamma_s^3}\right)^2 w^2 + \frac{\omega_{RF} q\hat{E}_{acc}}{cE_0\beta_s^3\gamma_s^3}\left[\sin(\phi_s + \varphi) - \varphi\cos\phi_s - \sin(\phi_s)\right] = \text{const} = H \tag{A5}$$

For each H we have different trajectories in the longitudinal phase space as schematically reported in Fig. A1.

The oscillations are stable within a region bounded by a special curve called separatrix whose equation is:

$$\frac{1}{2}\frac{\omega_{RF}}{cE_0\beta_s^3\gamma_s^3}w^2 + q\hat{E}_{acc}\left[\sin(\phi_s + \varphi) - (2\phi_s + \varphi)\cos\phi_s + \sin(\phi_s)\right] = 0 \tag{A6}$$

The region inside the separatrix is called rf bucket and the dimensions of the bucket shrink to zero if $\phi_s=0$.

Trajectories outside the rf bucket are unstable and the rf acceptance is defined as the maximum extension in phase and energy that we can accept in an accelerator. They are given by:

$$\Delta\varphi|_{MAX} \cong 3\phi_s$$

$$\Delta w|_{MAX} = \pm 2\left[\frac{qcE_o\beta_s^3\gamma_s^3\hat{E}_{acc}(\phi_s\cos\phi_s - \sin\phi_s)}{\omega_{RF}}\right]^{\frac{1}{2}}. \tag{A7}$$

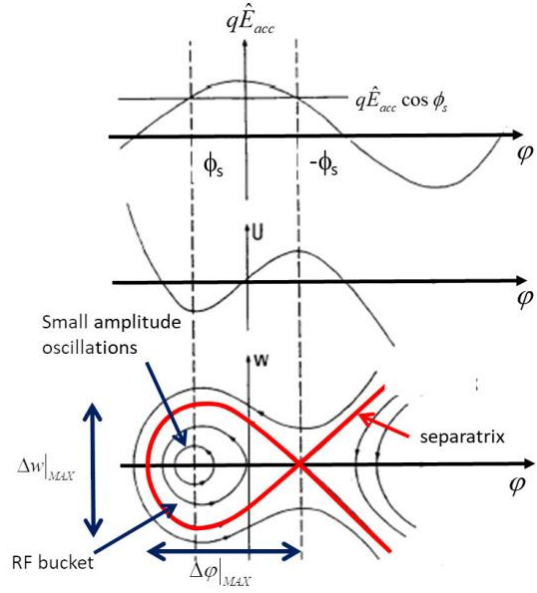

**Fig. A1:** Different trajectories in the longitudinal phase space corresponding to different values of H (the red one is the separatrix).